\documentstyle[12pt]{article}
\begin{document}

\newcommand{\nd}[1]{/\hspace{-0.6em} #1}
\begin{titlepage}
\begin{flushright}
CERN-TH.6309/91 \\
ACT-53\\
CTP-TAMU-90/91\\
\end{flushright}
\begin{centering}
\vspace{.1in}
{\large {\bf On the Evaporation of
Black Holes in String Theory}} \\
\vspace{.4in}
{\bf John Ellis} and {\bf N.E. Mavromatos}\\
\vspace{.05in}
Theory Division, CERN, CH-1211, Geneva 23, Switzerland. \\
and \\

\vspace{.05in}
{\bf D.V. Nanopoulos}\\

\vspace{.05in}
Center for Theoretical Physics, Dept. of Physics, \\
Texas A \& M University, College Station, TX 77843-4242, USA \\
\vspace{.03in}
and \\
\vspace{.05in}
Astroparticle Physics Group \\
Houston Advanced Research Center (HARC),\\
The Woodlands, TX 77381, USA\\
\vspace{.03in}
\vspace{.1in}
{\bf Abstract} \\
\vspace{.05in}
\end{centering}
{\small
\paragraph{}
We show that, in string theory, the quantum evaporation
and
decay
of black holes in two-dimensional target space is related
to imaginary parts in higher-genus string amplitudes.
These arise from the regularisation of modular infinities
due to the sum over world-sheet configurations, that are known
to express the instabilities of massive string states in general,
and are not thermal in character.
The absence of such imaginary parts
in the matrix-model limit confirms
that the latter
constitutes the final stage of the evaporation process,
at least in perturbation theory. Our arguments
appear to be quite generic, related only to the
summation over world-sheet surfaces, and hence should also
apply
to higher-dimensional target spaces. }
\par
\vspace{0.2in}
\begin{flushleft}
CERN-TH.6309/91 \\
ACT-53 \\
CTP-TAMU-90/91 \\
November  1991 \\
\end{flushleft}
\end{titlepage}
\paragraph{}
\newpage
\section{Introduction and Summary}
\paragraph{}
Recently a lot of attention has been paid to
black-hole solutions of two-dimensional string theories
\cite{gupt,witt,verl}, which are
discretized by
certain matrix models describing
$c=1$ conformal matter coupled to the
Liouville mode. This connection is
possible thanks to the discovery
of non-critical strings \cite{aben,myers},
that has opened the way for studying
non-trivial string dynamics.
These string theories are described
by non-compact coset
Wess-Zumino models on an arbitrary world-sheet \cite{witt,aben}.
They provide a powerful theoretical laboratory for seeing how ideas
about black hole physics fare in the context of string theory.
In particular, the existence of an infinite set of `W-hair' sufficient
to maintain quantum coherence has been conjectured and demonstrated
\cite{emn1,emn2}.
Several authors
\cite{vari} have now
extended the
black-hole
construction to higher-dimensional
target spaces by gauging more complicated groups,offering eventually
the hope of understanding black holes in four-dimensional target space.
\paragraph{}
So far all known
black-hole solutions are
{\it static}, which restricts their  physical
significance.
One expects that physical black holes are produced by
collapsing matter, and that quantum effects make them
evaporate \cite{hawk,birel}.
Therefore,
static solutions
cannot be the whole
story. The main
purpose of this note
to argue that the static character of the black hole is
a feature only
of the classical string tree
level,
and that formulating
the theory on {\it summed-up} higher-genus world-sheets leads to
quantum
instabilities, that correspond to
evaporation and decay
of the black hole, and do not have a
finite-temperature interpretation.
These instabilities manifest themselves as imaginary
parts in
string correlation functions arising from the
regularisation of modular
infinities \cite{marc}. These imaginary parts
are absent for the discrete states in the corresponding
$c=1$ matrix model, confirming (at least within string perturbation
theory) their interpretation
as the end-points of black hole evaporation, and strengthening
their role as guardian angels of quantum coherence \cite{emn1,emn2}.
We shall concentrate on the two-dimensional
black-hole case although our arguments appear to be quite generic
to
the divergences emerging from world-surface summations,
and can be applied to higher-dimensional cases as well.
\paragraph{}
The outline of the paper is as follows: in section 2 we review briefly
the origin
of the imaginary parts in correlation functions
in ordinary string theories.
In section 3 we
discuss some aspects
of the two-dimensional strings that will be
useful in our discussion. Section
4 is devoted to a discussion of
string propagation
in a Minkowskian
black-hole background on the torus.
We construct the corrections to
the tree-level effective action coming from summing up genus-zero and
-one
world-sheet surfaces, and show explicitly the existence of imaginary
mass-shifts for the black-hole solution. The latter are interpreted as
a signal for evaporation induced by quantum effcts in target space-time.
In section 5 we discuss
the interpretation of the Euclidean (thermal) black hole,
and
in section 6 we briefly discuss
higher-dimensional target-space black holes.
Finally, we present some conclusions and discuss
prospects for future progress in section 7.
\section{Critical String Theory in Higher Genera}
\paragraph{}
There is
a rather extensive literature on this issue
\cite{tsey,mizrachi,abe,fisch}.
Here we shall recall only the parts that are
relevant for our purposes, namely the regularisation
of modular infinities arising from
summing
over world-sheet
tori in the
one-string-loop approximation. The relevant analysis
has been done first by Marcus \cite{marc}, whose method will
be followed here.
\paragraph{}
Consider an N-point tachyon amplitude
in closed bosonic string theory. Unitarity requires
factorisation in the sense that the amplitude
exhibits poles whenever a particular combination
of external momenta approaches an `on-shell' value
for
any of the intermediate string modes. The formal origin of
the amplitude poles can be traced back to the well-known operator
product singularity, which occurs when two (or more) of
the vertex operators approach each other.
\paragraph{}
Using the factorisation formula,
\begin{equation}
  \Gamma(k_{1},...,k_{n})     \rightarrow
\sum_{states} \Gamma_{L}(k_{1},...,k_{l},-k)
\frac{1}{k^2 + m^2 -i\epsilon}
\Gamma_{R} (k,k_{l+1},...k_{n})
\label{fact}
\end{equation}

\noindent and combining the tree and torus amplitudes,
one can deduce
a mass renormalisation for the intermediate massive string modes,
which was discussed
by Weinberg \cite{wein}.
The relevant mass shift is given by the two-point function
on the torus:
\begin{equation}
     \delta m^{2} = - \Gamma (k,-k)
\label{twop}
\end{equation}

\noindent The latter expression in general diverges
when one sums over tori, as required in a string theory formulation.
Such a procedure involves integration over the
Teichmuller parameters
$\tau=\tau_{1} + i \tau_{2}$
which describe the various tori
(e.g. $\tau_{2}$ is related to the area of the torus). The modular
infinities arise from the region of the
$\tau_{2}$ integration
where $\tau_{2} \rightarrow \infty$, i.e. large tori,
and
are, therefore,
considered by many
as `infrared' infinities. As Marcus
\cite{marc} observed, the relevant part of the divergences
encountered in generic string amplitudes is similar in structure
to the `tachyon' infinity
of the
bosonic string. This means that even in
the superstring case, where the tachyons are absent,
there will be infinities of the form
\begin{equation}
  \int_{M>R} \frac{d^2 \tau}{\tau_{2}^{2}} \tau_{2}^{-12}
\frac{1}{|\Delta_{12}|^{2}}
\label{modinf}
\end{equation}

\noindent where the integration is over the part of the
fundamental region which is above the cut-off $R$. The final
results expressing regularised divergences turn out, as expected,
to be independent of $R$. The rest of the $\tau$-integration
over $M'<R$ yields modular-convergent results that we shall not
be concerned with in this note. The quantity $\Delta_{12}$
is a cusp form of weight $12$, related to the Dedekind $\eta$-function
by
$\eta(\tau)=q^{\frac{1}{24}}\Pi_{n=1}^{\infty} (1-q^{n}) \equiv
\Delta_{12}^{\frac{1}{24}}$, in standard
notation with $q=exp(i2\pi \tau)$. In the limit where
$\tau_{2} \rightarrow
\infty$, $\Delta_{12} \rightarrow exp(-2\pi\tau_{2})$. It is therefore
evident that the modular-divergent part has the generic form
\begin{equation}
  \int_{M>R} \frac{d^2 \tau}{\tau_{2}^{2}} \tau_{2}^{-12}
  e^{4\pi\tau_{2}}
\label{tau}
\end{equation}

\noindent It should be stressed again that this type
of divergence,
although it is similar in form
to the ones induced by tachyons in bosonic strings,
exists independently of tachyons in the theory.
This will be relevant for the case of two-dimensional
strings, where the tachyons are {\it massless}. Divergences
of the form (\ref{tau})
exist in that case as well.
\paragraph{}
Marcus \cite{marc} applied analytic continuation
to
regularise these modular divergences.
He used the integral
\begin{equation}
    \int_{R}^{\infty} \frac{dx}{x} x^{-\beta}  e^{-\alpha x}
\label{integr}
\end{equation}
\noindent which was first evaluated at $\alpha > 0$, and then he
analytically continued it to complex values of $\alpha$. The result is,
\begin{equation}
\int_{R}^{\infty} \frac{dx}{x} x^{-\beta} e^{(\alpha + i\epsilon)x}
=[\alpha e^{-i(\pi-\epsilon)}]^{\beta}
\int_{-\alpha R-i\epsilon}^{\infty} \frac{dx}{x} x^{-\beta} e^{-x}
\label{imp}
\end{equation}

\noindent The imaginary part of the integral is independent
of $R$, as expected, and constitutes the only remnant
of the divergence \cite{marc}:
\begin{equation}
   Im \int_{R}^{\infty} \frac{dx}{x} x^{-\beta}  e^{(\alpha+i\epsilon)x}
= \frac{\pi}{\Gamma(1+\beta )}\alpha^{\beta}
\label{inst}
\end{equation}

\noindent In this way , in ordinary string theories
one evaluates the imaginary part
of the one-loop induced
cosmological constant of the bosonic string, expressing the
{\it instability}
of the false tachyonic vacuum.
\paragraph{}
Imaginary parts,
similar to the one appearing in the cosmological
constant, also appear in higher-point functions of massive string modes,
and
reflect decay of the massive states with life-times determined
by $\Gamma=-Im(\delta m)$, where $\delta m$ is the mass
shift of the state in question.
\paragraph{}
In closed bosonic strings, imaginary parts also appear
in the two-
and three-point functions of the (massless)
graviton-dilaton multiplet, whose mass is now shifted
due to the induced cosmological constant term.
This graviton mass shift is consistent with general covariance
\cite{grav}.  Such lowest-order computations of string
amplitudes contain information about the string effective action,
to lowest order in target-space derivatives, or equivalently to
first order in the Regge slope parameter $\alpha '$.
The one-loop amplitudes, involving an
integration over tori,
yield in this way a cosmological constant term and a shift
in the Einstein term of the effective string action
\cite{tsey,abe}.
Standard computational techniques yield for the one-string-loop
corrected
effective action
the following form:
\begin{eqnarray}
\nonumber
   S_{eff}= \int d^D x \sqrt{G} e^{\Phi} [(1-\frac{3}{4\pi}
C_{torus}
  &\int_{M}&
    d^2 \tau \tau_{2}^{-14}
\frac{1}{|\Delta_{12}|^{2}}E(\tau,1))R + \\
C_{torus}
&\int_{M}&d^2 \tau
\tau_{2}^{-14} \frac{1}{|\Delta_{12}|^{2}} +... ]
\label{eff}
\end{eqnarray}

\noindent where $C_{torus}$ is a positive constant \cite{abe}.
The modular $\tau$-integration is over the fundamental
region and the function $E(\tau,1)$ is given by the formula \cite{abe}
\begin{equation}
  E(\tau,1)=lim_{\varepsilon
   \rightarrow 0}\frac{1}{\varepsilon} + 2\pi (\gamma -
\frac{1}{2}ln\tau_{2} -\frac{1}{12} ln|\Delta_{12}|)
\label{epsf}
\end{equation}

\noindent in
$\zeta$-function regularisation,
where $\gamma$ is Euler's constant. The logarithmic
$\frac{1}{\varepsilon}$ divergence in (\ref{epsf}) is `absorbed' in the
dilaton tadpole,
and corresponds to higher-genus corrections
to tree-level beta-function equations (local infinities)
\cite{fisch}. This is a rather general rule \cite{marc}.
Logarithmic divergences are the ones that
cannot be regularised by analytic continuation.
In addition to these divergences there are  $\tau_{2}$
modular
infinities arising from the region of $\tau_{2}$ integration
where $\tau_{2} \rightarrow \infty $.The latter are
the types of divergences that are going to interest us in this note.
They can be regularised by analytic continuation in the way
outlined in (\ref{integr}), yielding imaginary parts of the form
(\ref{inst}). Indeed,
the leading $\tau_{2}$ divergences in
the cosmological constant and Einstein parts of the action
(\ref{eff}) are
similar
in form,
up to irrelevant proportionality
constants,\footnote{The leading divergence in the $E(\tau,1)$ function
comes from $-ln|\Delta_{12}|$  which behaves like $\tau_2$ for large
$\tau_2$.}
and are of the type (\ref{integr}).
The regulated expressions have imaginary parts which in ordinary
bosonic string theories are attributed to the tachyonic contributions
in the string tadpole
graphs. However, as Marcus \cite{marc}
noticed,
such divergences
exist also in superstring theories where they
express
the decay of massive string states. For instance,
N-point
superstring amplitudes involving massless modes as
external states
contain similar divergences expressing
the decay rates of the exchanged states.
\section{Aspects of Two-dimensional Bosonic Strings}
\paragraph{}
The
consistent formulation of closed bosonic string theory in two
target-space dimensions is possible, provided one allows for
non-trivial backgrounds. It is sufficient to have a non-zero
dilaton background which is linear in the `spatial' target space
coordinate,
associated with the Liouville mode \cite{aben,myers,polch}.
The only propagating field in this string theory is the scalar
`tachyon'
mode - which is
however
massless in two dimensions.
The rest of the string modes are `topological' in the sense that
they make non-trivial contributions to amplitudes only for
particular values of the
energy and momenta \cite{poly}.
There is a simple reason for this. Consider,
for example,
the first `massive' string multiplet that consists of the
graviton-dilaton modes.
The massive character of the
multiplet is due to
the cosmological constant
terms in the action, as a result of the non-critical dimensionality
of space-time.
The graviton is massive with mass proportional
to $Q=\sqrt{\frac{25-c}{3}}=2\sqrt{2}$ $(c=1)$.
The conformal invariance conditions on a flat space
read \cite{pol,emn1}
\begin{eqnarray}
\nonumber
q^{\mu}(q_{\mu}+Q_{\mu}) &=&0 \\
(q^{\mu} + Q^{\mu})h_{\mu\nu}(q)&=&0\\
\nonumber q^{\mu} {\tilde h_{\mu\nu}}(q+Q)&=&0
\label{grav}
\end{eqnarray}

\noindent where $h_{\mu\nu}$ is the polarisation tensor for the
graviton-dilaton multiplet, $Q^{\mu}=(Q,0)$
and
the tilde denotes $Q$-conjugation in the sense of
\cite{poly} \footnote{In the original works on stringy representations
of Liouville theory \cite{seib},
people gave arguments for
disregarding $Q$-conjugate states. However,
it turns out that these
states have physical significance. For instance the $Q$-gravitons
constitute the last stage of black hole evaporation \cite{emn1,gupt},
as we discuss below.}.
In two-dimensional
target space,
the above relations
imply the decoupling of all the gravitons whose momenta are different
from $0$ or $-Q$,since they are longitudinal
and hence can be gauged  away.
However,
for the discrete values $0$ or $-Q$ there is a discontinuity
in the degrees of freedom, and extra modes
become relevant.
Due to the definite energy and momentum they possess,
their propagators
cannot be defined, since the latter
involve
analytic continuations
off-mass-shell. It is in this sense
that these modes are considered `semi-topological' or `of
co-dimension two' \cite{poly}.  The same is true for all the
higher string modes. Each one of them is associated with a
stringy {\it gauge} symmetry,
leading to
Ward identities satisified by the relevant string amplitudes
defined on-shell \cite{ven,emn1} \footnote{In this sense
two-dimensional target-space
general covariance is considered as the gauge symmetry associated
with the first excited string multiplet (graviton-dilaton).
In the two-dimensional example, general covariance
is
expressed via Ward-identities of the form \cite{ven,emn1}\\
$(q^{\mu}+Q^{\mu})<V_{\mu\nu}^{G}(q) \Pi_{j=1,...,N} V^{T}(k_{j})>=0$,
for a closed string amplitude involving, say, one graviton and N
tachyons.}.
Such Ward identities imply the gauging away of any massive
mode
whose momentum is off mass-shell, and this is the reason
why in two dimensions the only remnants of the higher
string states are {\it discrete} semi-topological modes
\footnote{It should be noticed that the presence of these extra modes
is necessitated by the
same arguments of
unitarity that require the  factorisation
of the convnentional $S$-matrix
in string theory.
A similar factorisation, but now involving the
exchange of co-dimension two states,
occurs in the
two-dimensional string case \cite{poly,sak},
thereby
explaining the extra poles in tachyon
scattering amplitudes of the $c=1$ theory \cite{poly}.}.
\paragraph{}
The impossibility of applying analytic continuation
to the mass-shell of the discrete massive states of the two-dimensional
string
implies the absence of any modular infinities
in closed string loops associated with these modes.
If this were not the case,
then according to the
Marcus analysis \cite{marc}
it would be impossible
to regulate such divergences
either by
absorbing them in tree-level $\sigma$-model coupling constants,
or by
analytic continuation, and therefore the theory
would be sick. Fortunately this is not the case. It
can be shown,
when
considering tachyon amplitudes on the torus, that
the only propagating field in the loop is again the tachyon
whilst
the massive states yield non-zero but finite contributions
\cite{poly}. The latter result has also been confirmed
in the
Das-Jevicki \cite{das}
string field theory approach
to the $c=1$ matrix model,
where it has been shown that
the extra discrete states yield non-zero but {\it real}
contributions to the scattering amplitudes of the massless
propagating fields (tachyons) of the theory \cite{dem}.
\section{Minkowskian Two-dimensional
Black Holes and the
Summation over Riemann Surfaces}
\paragraph{}
The previous Ward-identity
arguments about the decoupling of higher
string states
except at
{\it discrete} values of their momenta do not
apply
if some of the modes'
polarisation tensors
exhibit
{\it singularities}, as is the case of target space-time
black holes \cite{gupt,witt}. In that case, the familiar
Einstein terms in the effective action for graviton-dilaton
modes
appear,
and one finds the black-hole solutions
by the usual variational principles that one applies to
dynamical graviton fields in ordinary point-like
theories.
It will be useful,
for subsequent purposes,
to recall
some of the basic properties of the static black hole solutions
in 2D string theory.
\paragraph{}
Such objects are found as solutions of the beta function
equations for a bosonic $\sigma$-model to lowest order
in the Regge slope parameter $\alpha'$. It was Witten's
observation \cite{witt} that such  constructions
on an
arbitrary Riemann surface
arise from
appropriate gauging
of a Wess-Zumino coset model \footnote{The Minkowskian
black hole is obtained by gauging a non-compact
subgroup of
$SL(2,R)$.}
on $\frac{SL(2,R)}{U(1)}$
with the correspondence of $k-2$ to
$\frac{1}{4\pi
\alpha'}$,
where $k$
is the level
parameter
of the
Wess-Zumino term.The
lowest order (in $\alpha'$)
solutions
of the $\sigma$-model
correspond therefore
to the large-$k$ limit
of the group-theoretic model.
The effective action at tree level for graviton-dilaton
backgrounds reads
\begin{equation}
  \int d^2 x \sqrt{G}e^{\phi}(R + \Lambda+ O[(\nabla \phi)^{2}]+...)
\label{bh}
\end{equation}

\noindent where for our discussion we ignore matter (tachyon) parts.
In two-dimensional target spaces the non-trivial solutions
of the
equations
of motion obtained from (\ref{bh})
are of black-hole type \cite{gupt,witt},
leading to singular metrics in a certain coordinate system,
although, as usual, such singularities can be eliminated
by going to an appropriate coordinate system.
The black hole solutions
are static and can be thought of as the classical
final state of
gravitational collapse of two-dimensional matter \cite{mann}.
The mass of the black hole is essentially determined by
an arbitrary constant $\alpha$,
which expresses a shift
in the dilaton field. Indeed,
if we make the change
$\phi \rightarrow \phi+\alpha$,
the mass of the black hole
turns out to be \cite{witt}
$M_{bh}=\sqrt{\frac{2}{k-2}}e^{\alpha}$.
The family of black hole solutions
leads to
invariant line
elements
in space-time
given by the following expression \cite{gupt}
\begin{equation}
ds^{2} = (1-Me^{Q\rho})dt^2 - \frac{1}{1-Me^{Q\rho}}d\rho^{2}
\label{gupta}
\end{equation}

\noindent where $\rho$ is the space coordinate (Liouville mode).
In the limiting
case $M \rightarrow 0$,
the black hole solution (\ref{gupta}) becomes identical to a
Q-graviton perturbation, which can,
at least
in this sense, be identified
with the last stage of black hole evaporation.
In addition to the graviton mode,
all the rest of the higher massive string modes
are also excited. This underpins
the interpretation of
the continuum version of the $c=1$ matrix
model as
the final stage of the black-hole evaporation,
as
conjectured by Witten \cite{witt}.
\paragraph{}
It is the purpose of this section to discuss the origin of the
evaporation proccess. In ordinary local gravity theories
the evaporation of a black hole is a {\it quantum}
phenomenon which must therefore be
associated with
loop corrections
\cite{hawk,birel}. If a similar
mechanism operates in our case,
one should expect to see the evaporation process
when one performs the sum over world-sheet genera, that represents
the string analogue of the loop corrections to the gravity
action.
This is precisely what happens in our
case,
as we shall argue below.
We shall demonstrate our arguments by restricting
ourselves to the torus case, which is sufficient
for our purpose.
\paragraph{}
There is also
a formal reason for this.
In a stringy formulation of two-dimensional
quantum gravity,
the Liouville field is usually considered as
a free field
whose space is unrestricted. However, since this
mode is associated with the covariant short-distance cut-off
in the theory, it is natural to think of it as being bounded
from below at a value defining the cut-off, $\alpha$, in a flat
two-dimensional world-sheet
\cite{mir}. Representing
the covariant cut-off, then, as $e^{\rho}\alpha$,
the $\rho$ integration
extends from $0$ to $\infty$ \cite{pol}. These `boundaries' in Liouville
space have important consequences for
Liouville energy
non-conservation in string amplitudes, except in the torus case
\cite{pol}. To see this in a simplified way, let us complexify
the Liouville field by going to the
$i\rho$-formalism, and therefore
considering complex two-dimensional surfaces. Due to boundaries
in the integration over the zero-modes of $\rho$,
the result in string
amplitudes is not a delta-function conservation of the energy
associated with the coordinate $\rho$ (in a Fourier expansion of the
backgrounds) but rather resonant forms $\frac{1}{s}$ \cite{poly},
where $s=\sum_{i} \varepsilon (k_{i})      + Q (g-1)$, with
$\beta^{\mu} \equiv (\epsilon(k),k)$ being
the two-vectors representing
conformal charges in the
Liouville and
matter sectors, and the sum is over
states in the relevant string amplitude. The residues of these
resonances are the string
amplitudes we are considering \cite{poly}.
In any other topology
except that
of the torus,
the  Liouville energy conservation
law is modified by the `charge at infinity' $Q$
\cite{aben,poly}.
At genus one
one
there is
{\it exact} energy conservation
despite the presence of $Q$.
This makes the contribution of
this particular topology
somewhat special. This also implies that
the regularisation of the associated modular infinities coming
from this
topology
could not be cancelled by higher genera.
\paragraph{}
After these parenthetic remarks we are now in a
position to
discuss
the evaporation of the static black hole solution
(\ref{gupta}) induced by quantum effects
in
the torus case.
Energy conservation, even in the
in Liouville sector,
implies that the torus computation
can be considered formally identical to the one in critical
strings outlined in previous sections.
The only point that deserves attention concerns the
role of the target-space dimensionality. In the
case of closed bosonic
strings living in non-critical dimensions $D$
of target space-time the (modular invariant)
torus partition function is given by \cite{aben}
\begin{equation}
 Z_{D}=(2\tau_2)^{\frac{2-D}{2}}
 (\eta(\tau){\overline \eta(\tau)})^{2-D}
\label{twodim}
\end{equation}

\noindent Naively one expects no modular $\tau_2$ infinities
in $D=2$, and indeed this is the case in matrix model backgrounds
\cite{bersh}. However, when
computing correlation functions in Liouville
theory
with a world-sheet cosmological constant it seems
necessary
to continue
analytically
the matter central charge \cite{goulian},
which in turn implies a
formal continuation away from the $D=2$ value.
Upon such a procedure, which could be viewed as
the analogue of target-space
dimensional regularisation, loop corrections to the string
effective action
acquire,
in the limit $D \rightarrow 2^{+}$,
{\it finite} imaginary parts from the regularisation
of $\tau_2$-modular infinities that appear
in the case
$D > 2$,
as becomes clear from (\ref{twodim}).
Similarly
to the critical
string theory,
the imaginary parts of the
corrections to the Einstein term
of the two-dimensional string
are given by
the $\alpha \rightarrow 0$ and $\beta \rightarrow 0$
limit of (\ref{inst}): the result is $\pi$.
The imaginary parts of the torus correction to the (tree-level)
cosmological
constant, on the other hand,
are given by (\ref{inst}) upon setting
$\alpha=\varepsilon \rightarrow 0$, $\beta=1$: the
result is
$\varepsilon \frac{\pi}{2}$ \footnote{It should be
noted that this is the form
of the imaginary parts in the correction to the Einstein term
arising from regularisation of subleading $\tau_2$ divergences.
If there are singularities in the curvature terms, as
is the case of the Minkowskian
black hole solution in the asymptotic region
where the dilaton approaches $-\infty$ \cite{witt},
then the above analysis implies
the existence
of additional
{\it finite} imaginary parts in the torus correction
to the Einstein term in the effective action.}. Hence,
in the case of two-dimensional strings there are no
imaginary parts in the one-string-loop corrected
cosmological constant. This reflects
the stability of the two-dimensional flat-space
tachyonic vacuum, which is
{\it massless}.

\paragraph{}
The important point is
that
the proportionality constant in front of
the Einstein term can be identified with
the mass of the black hole \cite{witt}. If
$a$ represents a constant shift in the dilaton, then, by
computing the stress-tensor of
the graviton-dilaton system (in the absence of matter)
one can determine the (conserved) energy, i.e. mass,
of the black hole as
\begin{equation}
M_{bh} = \sqrt{\frac{2}{k-2}}e^{a}
\label{mass}
\end{equation}

\noindent Therefore,
the torus contribution is to shift the mass of the
tree-level black hole by an amount related to an
infinite integral over the
moduli space of genus-one surfaces. The latter,
as we have
mentioned,
has both logarithmic divergences, that can be
absorbed in a renormalisation of the dilaton field at tree-level
\`a la Fischler and Susskind \cite{fisch}, and modular $\tau_{2}$
infinities whose analytic continuation and
regularisation
yield imaginary parts in the black-hole mass
shifts \footnote{In string perturbation theory the real
part
of the one-string-loop
correction
to the
Einstein-term
cannot reverse the positive sign of the tree-level coefficient,
so the combined result of the tree and one-loop string level
computations
can still be represented as
an exponential of a shifted dilaton
field.}. The situation, therefore, is similar in nature to what
happens in critical string theories, where massive string
states acquire complex mass-shifts in higher genera, reflecting
their decay with a life-time inversely
proportional in  magnitude to the
imaginary part of the pertinent mass-shift.
In our case it is the Minkowski
black-hole state that is {\it unstable}
due to quantum effects, although classically, at the
tree string-level,
it is a stable background configuration.
It is in this sense
that we exhibit the evaporation of the two-dimensional quantum
black hole. In this point of view, evaporation is expressed
as an {\it instability} of the black hole vacuum with respect to
stringy quantum corrections.
It should be stressed
that
imaginary parts arising from the regularisation
of modular $\tau_2$-infinities appear
in any
target-space
dimensionality  $D \ge 2$, which gives a sort of
universality to this decay
mechanism.
This mechanism for evaporation is purely stringy
and has no counterpart in local gravity theories. The
imaginary parts that
express
the instability
arise from
the
regularisation of large-area tori,
and therefore rely on
the concept of an underlying world-sheet structure, i.e. string
theory. This is to be contrasted with
local point-like theories of
gravity, where such phenomena do not occur. One could still
refer to this type of process as `Hawking radiation',
due to the fact that it is triggered by {\it quantum} effects,
but is not thermal
as in the local field theory case.We cannot, however,
yet
exclude the possibility that
thermal instabilities might
arise in our picture
in higher-dimensional target-spaces.
\paragraph{}
We can now answer questions concerning the manner of the
black hole
evaporation,
as well as its final stage.
As argued in \cite{emn1,emn2} there is an infinity of
gauge conservation laws that accompany black hole solutions
in two-dimensional strings, which express stringy gauge
symmetries associated with the infinite tower of string states.
These laws imply the existence of {\it conserved} charges
that constitute  the `hair' of black holes.
These quantum numbers,
being expressible as total
spatial derivatives, remain conserved during the evaporation
process, thereby restricting the modes of black-hole
radiation.
It can be shown
that the gauge group associated with these charges contains
classical $w_{\infty}$-symmetries \cite{mooreseib} which
preserve
the phase-space area (two-dimensional
volume)
of the matrix model \cite{witt2}.
There
is no loss of quantum coherence due to the evaporation process,
for
the
reasons explained in \cite{emn2}.
{}From the observation that the vanishing mass limit
of the
black-hole solution
describes the
$Q$-graviton
and the other discrete
topological states of the two-dimensional string, as well as
the fact that the contribution of the latter to scattering amplitudes
determining the matrix-model target-space effective action contains
{\it no imaginary } parts, we conclude that the continuum
version of the $c=1$ matrix model constitutes, at least in
string perturbation theory,
the final stage of
black-hole evaporation.
\paragraph{}
Unfortunately,
it is not known how to  perform
in the continuum language
the sum over genera in closed string cases \cite{peri}.
The matrix model approach for $c<1$ looks helpful,
but the situation concerning
$c=1$ matrix models is still unclear.
However,
in our case we have shown that higher-genus
corrections make
black hole solutions unstable
{\it even in perturbation theory}, thereby implying their decay
(evaporation). At least as far as two-dimensional continuum string
theory is concerned,
the perturbation theory result seems to indicate
that the flat-space linear dilaton background solution of the latter
is the final point of the evaporation. It is of course possible
that non-perturbative effects
lead to
a different end-point, but such a possibility goes beyond the
scope of this analysis.
\paragraph{}
A final comment we would like to make in this section
concerns the possibility of regarding the
static (classical) black hole solution as the result
of some sort of gravitational collapse.
This would be useful in considering the two-dimensional
black hole as a laboratory for the study of higher-dimensional
physically interesting cases, where such phenomena occur.
In
two dimensions the concept of collapsing
matter is not obvious. The matter-stress tensor obtained
naively from the
Einstein tensor vanishes,
since in two-dimensional
target spaces $R_{\mu\nu}-\frac{1}{2}G_{\mu\nu} R$ vanishes
identically. However,
in string effective theories there are non-trivial
dilaton terms that accompany the Einstein curvature term in the effective
action. Their presence make the matter (tachyon) stress tensor
non-zero
even classically \cite{emn1,dealwis}.
Shifting the dilaton field
by a constant
defines
a family of objects characterised by various masses.
Thus collapsing
matter could
lead to a black-hole. By matching -
in the boundary of the dust (matter) -
the static
black hole solutions
studied in \cite{witt,verl},
with the solutions obtained from non-zero matter-stress tensors,
it can be shown \cite{mann} that the former
correspond
to the final state of such a collapse,
at the
classical level.
\section{Euclidean Black Holes}
\paragraph{}
Euclidean black holes are described by coset models
in which the gauged subgroup of the $SL(2,R)$ is {\it compact}.
They can be thought of as being obtained from the euclideanisation
of the Minkowski time \cite{witt}. From the point of view
of Liouville theory,
Euclidean black holes correspond
to a two-dimensional Liouville theory coupled to matter
described by a field compactified on a circle \cite{bersh,kutas}.
In such models the  partition function on the torus
is known
to possess no
modular $\tau_{2}$ infinities, since the
pertinent $\eta$-function factors cancel between
Liouville and matter sectors \cite{bersh}.
In physical terms,
this means that
such black holes are static thermal
objects which are in constant interaction with a heat
bath, and therefore one does not expect them to lose any
mass \cite{verl}. We cannot yet
exclude the possibility that
in higher-dimensional target spaces,
there
might be thermal instabilities
of such objects in higher genera, due to the
non-cancellation of $\eta$-factors that generate
modular infinities. However, there is no
example known
so far that
supports this idea.
For instance,
in certain string
theories,
like the open $D=4$ string \cite{dunb}, it
is known that below the Hagedorn temperature,
although there exist
thermal
mass shifts which do not occur in the
zero-temperature formalism
\cite{thermal}, nevertheless
they are modular
finite. It might therefore be that there exists
some sort of cancellation of `infrared' infinities
in thermal (static) string configurations, at least in
certain cases \cite{le bellac}.
However such an analysis
falls beyond the scope of the present
work.
\paragraph{}
Before closing this section it is worth
mentioning another difference between
the Euclidean and the Minkowski framework
within the context of two-dimensional strings.
In models with a {\it compactified} matter boson field
(as is also
the case of $c<1$ matrix models) the form of the
Ward identities that express certain stringy symmetries
is modified with respect to the uncompactified case.
The higher string modes do not decouple any longer
due to the discrete matter momenta. Consider,
for instance, the graviton-dilaton Ward identity expressing general
covariance of the theory. In the compactified case it reads
\cite{ven,poly}
\begin{equation}
 (q^{\mu} + Q^{\mu})<V_{\mu\nu}^{G}(q) \Pi_{j=1,...,N} V^{T}(k_j)>=
 \sum_{j=1,...,N} k_{j\nu}
 <V^{T}(k_j+q)\Pi_{i \ne j}V^{T}(k_i)>
\label{discr}
\end{equation}

\noindent The right hand-side part is a contact term and
vanishes `on-shell'
for the uncompactified case due to analyticity properties
(the well-known \cite{green} `cancelled propagator argument').
This is no longer true in the compact case due to the discrete
momenta in the matter sector \cite{poly}.
This can be interpreted formally
as the
breakdown of general covariance which is to be expected
in a thermal theory. This
is one of the main
differences between Minkowskian and Euclidean formalisms
of the $c=1$ string theory.
\section{Higher-Dimensional Target Spaces}
\paragraph{}
Analogous descriptions
of higher-dimensional
singular space-times
seem
possible. Support for this was recently given
in a number of works \cite{vari,eguchi} where
it was shown that by gauging more complicated coset
Wess-Zumino models
(even supersymmetric ones \cite{eguchi}) on arbitrary
Riemann surfaces,
one can get
interesting singular
field configurations for higher dimensional
target spaces, among which one finds black holes \cite{hor}
and black strings \cite{horow}.
In view of the arguments presented in this work, the
summation over Riemann surfaces is expected
to lead to instabilities
in the Minkowski
formalism. That this is indeed the case
can be demonstrated explicitly by looking at the
four-dimensional black hole of ref. \cite{hor} and the
three-dimensional black string of \cite{horow}.
Let us start from the latter.
\paragraph{}
The existence of the antisymmetric tensor field $B_{\mu\nu}$
(which is gauged away in two dimensions) leads to
a conventional {\it axionic} charge $Q$ for the black hole,
in addition to its mass $M$.
For completeness,
we briefly outline the construction \cite{horow}.
One adds a free boson field $z$ to the Wess-Zumino action,
which is equivalent to considering a
group
$G=SL(2,R) \times R$.
Then one gauges an appropriate one-dimensional subgroup generated
by that of the two-dimensional case \cite{witt} together with a
translation in the boson field $z$. This  leads to two arbitrary
parameters in the problem, the Wess-Zumino
level parameter $k$,
and an
extra parameter, $\lambda$, which is
associated with the translation of $z$.
The effective action
describing gravitational dynamics contains again
a non-trivial dilaton conformal factor acompanying
the Einstein term. In non-singular curvature cases
in dimensions higher than two,
one can
absorb these
factors in a conformal rescaling of the graviton.
However,
we choose not to do it in singular
cases
such
as black strings.
The reason is that the arbitrariness
in shifting the dilaton field by a constant $\alpha$
determines a family of solutions characterised
by various values of $Q$ and $M$ \cite{horow},
\begin{eqnarray}
\nonumber
    Q& = & e^{\alpha} \sqrt{\frac{2\lambda (1+\lambda)}{k}}\\
    M& = & e^{\alpha} (1+\lambda) \sqrt{\frac{2}{k}}
\label{axi}
\end{eqnarray}

\noindent The
summation over higher
genera will produce
imaginary parts  in $\alpha$ and therefore one has complex
shifts for the
black hole mass and axion charge,
which are again
interpreted as signal for evaporation. The {\it quantum}
three-dimensional black string
will therefore
evaporate parts of its mass
and axion charge.
The exact conservation of
these quantum numbers (due to their being total
space derivatives) implies that
the evaporated parts of the charges have to be carried away
by particles emitted from the black hole.
\paragraph{}
In a similar way,
one can study four-dimensional
black hole solutions \cite{hor}.
These may be
obtained by twisted
products of one Euclidean and one
Minkowskian two-dimensional black hole.
The antisymmetric tensor vanishes in this case, but
the four-dimensional metric is off-diagonal.
Again the role of the dilaton field is essential
in defining families of solutions, and our arguments
on the
instabilities induced by higher genera apply as in the black string
case.
One can even construct direct higher-dimensional black holes
by gauging more complicated Wess-Zumino models, involving
antisymmetric tensor fields \cite{quev}. In such a case
there is axionic charge on the black hole solutions.
{}From standard field-theoretic arguments \cite{gid}
one expects the three-dimensional
black hole not to evaporate
all of its axion
charge.
Similarly
to the two-dimensional
case,
however, one expects
quantum coherence to be restored,
not because of the axion charge alone,
but because of an infinity
of hair provided by the higher string modes. In two dimensions
this type of hair is phase-space area-preserving \cite{emn2,witt2}.
In view of the general description of
black holes by Wess-Zumino models,
and the relation of the latter
in certain cases to
such area-preserving ($w$-type) symmetries
\cite{park} we
conjecture that the existence of a
phase-space volume-element-preserving
symmetry is a general
feature of such a
Wess-Zumino theory,
and thus coherence is maintained during the
evaporation
process.
This fits in with the observation
\cite{preskill}
that the Hawking temperature of an evaporating
black hole is reduced as the
amount
of hair characterising the black hole is
increased. This leads one to
expect
that
a black hole with
infinite hair
evaporates
at zero
temperature \footnote{This is
consistent with the fact that the two-dimensional
black hole under consideration
resembles an extreme Reissner-Nordstrom
type \cite{witt}.}.
However,
this conjecture remains to be investigated further
in future work.
\paragraph{}
Another
comment
concerns the
extension of these ideas to supersymmetric cases.
Supersymmetric Wess-Zumino models have been considered \cite{nois}
with the result that the effective two-dimensional
theory describes supersymmetric
$\sigma$-models in
black hole backgrounds.
In space-time
supersymmetric backgrounds there are non-renormalisation theorems
\cite{mizrachi} that prevent the Einstein terms from receiving
corrections in higher genera. This is
equivalent
to the vanishing of the dilaton tadpoles in supersymmetric string
theories. One might naively think that
the
breaking of space-time supersymmetry by the black hole background
seems to be essential in allowing
imaginary parts in the higher-genus
corrections to the
mass of the black hole, and hence leading to evaporation
in a way similar to the bosonic case. However, in view of Marcus's
analysis \cite{marc},
instabilities of the kind discussed in the present work
also appear in superstring theories, expressing in general
the decays of massive states. An explicit example of such a situation
is the type-I
open superstring
\cite{tsuch}.
Thus we expect that a
similar mechanism
of non-thermal
black hole decay will also operate in superstrings and heterotic
strings.
\section{Conclusions and Prospects}
\paragraph{}
Let us now summarise the view of black hole quantum physics
that has emerged from this analysis and our previous papers
\cite{emn1,emn2}. We have identified an infinite set
of gauge $w$-symmetries that are sufficient to maintain
quantum coherence for two-dimensional black holes \cite{emn1},
and characterize an infinite set of `topological states'
that constitute the final states of black hole evolution described
by a matrix model \cite{matrix}.
These $w$-symmetries have the geometrical interpretation
\cite{witt2} of preserving the phase-space  volume-element
of the matrix model, and thereby
exclude
\cite{emn2}
the general form of non-quantum-mechanical,
non-Hamiltonian modification of the evolution equation for the density
matrix \cite{ehns}, which would
otherwise have caused all quantum-mechanical systems to
appear `open' as a conjectured consequence of microscopic
space-time topology change \cite{hawk2}. In this paper we have
shown that the evaporation of two-dimensional Minkowskian black
holes can be understood as a quantum instability appearing in
higher genera, analogously to the normal decays of massive
string states \cite{marc}. This mechanism for Minkowskian
black hole decay does not have a direct thermal interpretation,
and, moreover,
the two-dimensional Euclidean finite-temperature black hole
solution \cite{verl} is {\it static} and does not exhibit
this decay instability \cite{bersh}.
\paragraph{}
It is now appropriate to speculate on the possible extension
of these results to four-dimensional black holes. As yet,
there is no general characterisation of four-dimensional stringy
black holes at the conformal field theory level, but some partial
results are becoming available \cite{vari}. These generalize, and
often incorporate, the non-compact coset Wess-Zumino models describing
the two-dimensional solutions. As such, we expect them to
include and extend the $w$-symmetries that save quantum coherence
in two dimensions. A very large set of gauge symmetries is in fact
known to
exist in generic four-dimensional string
models \cite{ven},
possibly in correspondence to the number of massive string states
and providing an amount of `hair' sufficient to quench
the entropy
usually associated with four-dimensional black holes
\footnote{For another approach, see \cite{kal}.}.
It certainly
seems that the intrinsically stringy non-thermal
higher-genus black hole
decay mechanism identified in this paper could carry over to
four dimensions, although we cannot yet exclude the existence
of additional thermal instabilities. It seems that the string
answer to the conundrum of reconciling quantum mechanics
with general relativity is at hand.
\paragraph{}
{\Large{\bf Acknowledgements}} \\
\par
We would like to acknowledge useful discussions with
J.L. Miramontes and J. Sanchez-Guill\`en.
The
work of D.V.N. is partially supported by DOE grant
DE-FG05-91-ER-40633.

\end{document}